\documentclass[a4paper,fleqn,usenatbib,useAMS]{mnras}

\usepackage[T1]{fontenc}
\usepackage{ae,aecompl}

\usepackage{graphicx}	
\usepackage{stackengine}
\usepackage{epstopdf}
\usepackage{savesym}
\usepackage{amsmath}
\savesymbol{iint}
\savesymbol{iiint}
\usepackage{txfonts}
\restoresymbol{TXF}{iint}
\restoresymbol{TXF}{iiint}
\usepackage{fix-cm}
\usepackage{subfig}
\usepackage[]{units}
\usepackage{textcomp}

\title[Non-axisymmetric instabilities in discs]{Non-axisymmetric instabilities in discs with imposed zonal flows}

\author[R. Vanon \& G.I. Ogilvie]{
R. Vanon\thanks{E-mail: \href{mailto:rv288@cam.ac.uk}{rv288@cam.ac.uk}}
and G. I. Ogilvie
\\
Department of Applied Mathematics and Theoretical Physics, University of Cambridge, Centre for Mathematical Sciences, Wilberforce Road, Cambridge\\
CB3 0WA, UK
}

\date{Accepted XXX. Received YYY; in original form ZZZ}

\pubyear{2016}

\begin{document}
\label{firstpage}
\pagerange{\pageref{firstpage}--\pageref{lastpage}}
\maketitle

\begin{abstract}
We conduct a linear stability calculation of an ideal Keplerian flow on which a sinusoidal zonal flow is imposed. The analysis uses the shearing sheet model and is carried out both in isothermal and adiabatic conditions, with and without self-gravity (SG). In the non-SG regime a structure in the potential vorticity (PV) leads to a non-axisymmetric Kelvin-Helmholtz (KH) instability; in the short-wavelength limit its growth rate agrees with the incompressible calculation by \citet{Lithwick2007}, which only considers perturbations elongated in the streamwise direction. The instability's strength is analysed as a function of the structure's properties, and zonal flows are found to be stable if their wavelength is $\gtrsim 8H$, where $H$ is the disc's scale height, regardless of the value of the adiabatic index $\gamma$. The non-axisymmetric KH instability can operate in Rayleigh-stable conditions, and it therefore represents the limiting factor to the structure's properties. Introducing SG triggers a second non-axisymmetric instability, which is found to be located around a PV maximum, while the KH instability is linked to a PV minimum, as expected. In the adiabatic regime, the same gravitational instability is detected even when the structure is present only in the entropy (not in the PV) and the instability spreads to weaker SG conditions as the entropy structure's amplitude is increased. This eventually yields a non-axisymmetric instability in the non-SG regime, albeit of weak strength, localised around an entropy maximum.
\end{abstract}

\begin{keywords}
accretion, accretion discs -- zonal flows -- instability -- turbulence
\end{keywords}



\section{Introduction}
Zonal flows are azimuthally symmetric shear flows exhibiting a strip-like alternating pattern which are observed in a wide range of settings.  
Possibly the most famous natural example is represented by Jupiter's belts, whose origin is still controversial; recent numerical simulations have confirmed and reproduced the presence of zonal flows in both gas and ice giant planets by modelling convective turbulence \citep[e.g.][]{Sunetal1993, HeimpelAurnou2007}.
Zonal flows also occur in laboratory plasma experiments, where they are generated by non-linear transfer of energy from drift waves between small and large scales \citep{Diamondetal2005}.

More recently, they have been encountered in 2D numerical simulations of shearing sheet models of accretion discs in magnetohydrodynamic (MHD) turbulent regimes \citep[e.g.][]{Johansenetal2009, Simonetal2012, KunzLesur2013, BaiStone2014}. They are seen to become more prominent and possess larger amplitudes with increasing box size, with their wavelength usually taking the largest possible value in the radial direction (i.e. $k_x = \frac{2\pi}{L_x}$, $k_x$ and $L_x$ being the wavenumber and the box size in the $x$ direction) \citep{Johansenetal2009, Simonetal2012}. The zonal flows are also observed to have longer lifetimes when considering larger box sizes \citep{BaiStone2014}, except when very small boxes are considered \citep{Johansenetal2009}, and they do not appear to be dependent on the initial conditions applied to the system, with \citet{Simonetal2012} initialising their simulations with either two flux tubes or a uniform toroidal field and seeing no difference in the wavenumber of the structure nor a significant discrepancy in its appearance time.

Using a 2D shearing sheet model of a hydrodynamical disc, \citet{Lithwick2007, Lithwick2009} showed that zonal flows in an incompressible, inviscid fluid can be unstable to the formation of vortices \citep[see also][]{Gill1965, LernerKnobloch1988} via the Kelvin-Helmholtz (KH) instability. The instability is however often missed by numerical simulations as it operates at small values of the azimuthal wavenumber $k_y$, implying that the computational domain must be substantially elongated in the $y$ direction.
Some 2D numerical simulations have shown that coherent vortices associated with zonal flows \citep[e.g.][]{UmurhanRegev2004, JohnsonGammie2005} appear to be long-lived; what's more, as \citet{UmurhanRegev2004} point out, accretion discs are characterised by Reynolds numbers $Re = VL/\nu$ ($V$ and $L$ being the characteristic velocity and length scales in the disc and $\nu$ the kinematic viscosity) that are orders of magnitude larger than what it is possible to implement in a simulation, which would potentially make vortices' lifetimes in discs extremely long. The vortices are also seen to produce an outward transport of angular momentum, necessary for accretion onto the central object to occur; this could thefore potentially prove to be an alternative to magnetohydrodynamic (MHD) turbulence in discs -- or regions of discs -- which are not sufficiently ionised for MHD turbulence to take place, as long as a suitable source for the zonal flow is pinpointed.

Lastly, zonal flows can play an important role in the dynamics of planetary formation in protoplanetary discs. A key obstacle in the early formation of protoplanets is the `metre-size barrier', representing the difficult chances of survival and growth of metre-sized bodies during the period of fast inward migration they experience due to gas drag \citep{Weidenschillingold1977}. The presence of density and velocity fluctuations in the disc, such as zonal flows, alters the drag force exerted on the material by the gas, potentially slowing down -- or halting altogether -- the inward drift of metre-sized bodies \citep{Whipple1972, KlahrLin2001, HaghighipourBoss2003, FromangNelson2005, Katoetal2009}. This would create a local enhancement in their density, promoting both the growth of these bodies through coagulation \citep{Weidenschilling1997, DullemondDominik2005, Braueretal2008a, Braueretal2008b} and the triggering of gravitational instability, potentially leading to fragmentation \citep{YoudinShu2002, Johansenetal2006, Johansenetal2007}.

Owing to the widespread interest in zonal flows mentioned above, the aim of this paper is to perform a linear analysis on the stability of a Keplerian accretion disc when an axisymmetric structure is imposed on the system; such analysis will be conducted using the compressible shearing sheet (2D) approximation in both isothermal and adiabatic conditions, also taking into account the effects of the disc's self-gravity (SG).
The paper is arranged as follows: in Section~\ref{sec:model} we present the basic equations of the shearing sheet model employed to tackle the problem, and derive the equations governing both the axisymmetric structures and their non-axisymmetric disturbances. We also describe the method by which they are solved numerically. In Section~\ref{sec:results} we present results on instabilities in both isothermal and adiabatic cases and compare these with relevant literature. The paper closes in Section~\ref{sec:conclusions} with the conclusions drawn from the results.

\section{Model}
\label{sec:model}
The analysis presented is based upon the unstratified shearing sheet model first employed by \citet{GoldreichLynden-Bell1965}. This is a local model which employs a Cartesian frame of reference which corotates with the Keplerian disc at some fiducial radius $R_0$ with angular frequency $\Omega \, \mathbf{e}_z$, where $\mathbf{e}_z$ is the unit vector parallel to the $z$-axis. The dimensions of the shearing sheet -- which is centred about the fiducial radius $R_0$ -- are chosen such that $L_x$, $L_y \ll R_0$, where $L_x$ and $L_y$ are the radial and azimuthal dimensions of the sheet.

In this frame of reference, the continuity equation and the momentum equation for an inviscid, compressible calculation can be written as

\begin{gather}
    \partial_{t} \Sigma + \nabla \cdot \left(\Sigma \boldsymbol\varv\right) = 0,\\ \label{eq:continuity}
  \partial_{t} \boldsymbol\varv + \boldsymbol\varv \cdot \nabla \boldsymbol\varv + 2 \boldsymbol\Omega \times \boldsymbol\varv = - \nabla \Phi - \nabla \Phi_{d,m} - \frac{1}{\Sigma} \nabla P, 
\end{gather}
where $\Sigma$ is the disc surface density, $\boldsymbol\varv$ is the flow velocity, $\boldsymbol\Omega$ the disc angular velocity, $\Phi = - q \Omega^2 x^2$ the effective tidal potential, $q = - \mathrm{d} \ln \Omega/\mathrm{d} \ln r$ the dimensionless rate of orbital shear, $\Phi_{d,m}$ the disc potential evaluated at its mid-plane and $P$ the 2D pressure.

The continuity equation can be expressed in terms of the quantity $h = \ln \Sigma + \mathrm{const.}$; this gives

\begin{equation}
  \frac{\partial h}{\partial t} + \boldsymbol\varv \cdot \nabla h + \nabla \cdot \boldsymbol\varv = 0.
\end{equation}

The expression for the disc potential is found by making use of the Fourier method to solve Poisson equation $\nabla^2 \Phi_d = 4 \pi G \Sigma \delta(z)$, where $\delta(z)$ represents Dirac's Delta function; this is most easily expressed in Fourier space as

\begin{equation} \label{eq:phi-d}
  \tilde{\Phi}_{d} = - \frac{2\pi G \tilde{\Sigma}}{\sqrt{k_{x}^2 + k_{y}^2}} \mathrm{e}^{-k\lvert z \rvert},
\end{equation} 
where $k_x $ and $k_y$ are the $x-$ and $y-$components of the wavevector $\mathbf{k} \neq 0$, $G$ is the gravitational constant and $z$ the height above the disc midplane. The above expression can then readily be evaluated at the disc midplane ($z=0$), resulting in

\begin{equation} \label{eq:phi-dm}
  \tilde{\Phi}_{d,m} = - \frac{2\pi G \tilde{\Sigma}}{\sqrt{k_{x}^2 + k_{y}^2}}.
\end{equation}

Two important quantities in our analysis are the potential vorticity $\zeta$ and the dimensionless specific entropy $s$, both of which are material invariants; these are defined as 

\begin{equation} \label{eq:PV}
  \zeta = \frac{2\Omega + (\nabla \times \boldsymbol\varv)_{z}}{\Sigma},
\end{equation}

\begin{equation}
  s = \frac{1}{\gamma} \ln P - \ln \Sigma,
\end{equation}
where $\gamma$ is the adiabatic index.

The basic state, described by $\boldsymbol\varv_0 = (0, -q \Omega x, 0)^{T}$ and $\Sigma = \Sigma_0$, is then perturbed such that $\boldsymbol\varv = \boldsymbol\varv_0 + \boldsymbol\varv^\prime$ (where $\boldsymbol\varv^\prime = (u^\prime , \varv^\prime , 0)^{T}$), $\Sigma = \Sigma_0 + \Sigma^\prime$, etc. The pressure term $\nabla P/\Sigma$ can be expressed in terms of $h$ and the internal energy per unit mass $e$ employing the relation $P = (\gamma-1) \Sigma e$; 
also, it is important to notice that the term $2\mathbf{\Omega} \times \boldsymbol\varv_0 = 2 \Omega^2 q x \mathbf{e}_x$ cancels out with the gradient of the effective potential $\nabla \Phi$.
This produces the following set of fully non-linear equations for the velocity, $h$ and internal energy per unit mass ($e$) perturbations, which are used in both isothermal and adiabatic conditions. 

\begin{equation} \label{eq:nonlin_h}
  \mathrm{D} h^{\prime} = - \left( \partial_x u^\prime + \partial_y \varv^\prime \right) \equiv - \Delta,
\end{equation}
\begin{equation} \label{eq:nonlin_u}
  \mathrm{D} u^{\prime} - 2\Omega \varv^\prime = - \partial_x \Phi_{d,m}^\prime - \frac{v_s^2}{\gamma}\partial_x h^\prime - (\gamma-1)\left[\partial_x e^\prime + e^\prime\partial_x h^\prime\right], 
\end{equation}
\begin{multline} \label{eq:nonlin_v}
  \mathrm{D} \varv^\prime + (2-q)\Omega u^\prime = - \partial_y \Phi_{d,m}^\prime - \frac{v_s^2}{\gamma}\partial_y h^\prime + \\ - (\gamma-1)\left[\partial_y e^\prime + e^\prime\partial_y h^\prime\right], 
\end{multline}
\begin{equation} \label{eq:nonlin_e}
  \mathrm{D} e^\prime = - (\gamma-1) \left(e_0 + e^\prime\right) \Delta, 
\end{equation}
where $\mathrm{D} = \partial_t + u^\prime \partial_x + (-q\Omega x+ \varv^\prime)\partial_y$ is the Lagrangian derivative, $e_0 = \nicefrac[]{\varv_s^2}{\gamma(\gamma-1)}$ and $\varv_s$ are the internal energy and sound speed of the background state and we have neglected any cooling and heating. 

In the linear approximation, the solution to these equations are shearing waves of the form $\Sigma^\prime(\mathbf{k}, \mathbf{x}) = \Re \left\{ \tilde{\Sigma}^{\prime}(t) \exp \left[ \mathrm{i} \mathbf{k}(t)\cdot\mathbf{x}\right]\right\}$ \citep{Kelvin1887}
where $\tilde{\Sigma}^{\prime}$ is the amplitude of the Fourier transform of $\Sigma$.
This solution causes the time-dependence of $\mathbf{k} = (q \Omega k_y t + \mathrm{const.}, \,\, \mathrm{const.})^{T}$, representing the action of the background shear flow on the wavefronts, to have repercussions on the shear terms $-q\Omega x \frac{\partial h^{\prime}}{\partial y}$, $-q\Omega x \frac{\partial u^{\prime}}{\partial y}$ and $-q\Omega x \frac{\partial \varv^{\prime}}{\partial y}$ (which originate from the $\boldsymbol\varv_0 \cdot \nabla h^{\prime}$ and $\boldsymbol\varv_0 \cdot \nabla \varv^{\prime}$ terms, respectively). This is because

\begin{align}
 (\partial_t - q \Omega x \partial_y) u^{\prime} = & \Re{ \left\{ \left[ \partial_t \tilde{u}^{\prime} + \mathrm{i} \left( \frac{\partial \mathbf{k}}{\partial t} \cdot \mathbf{r} - q \Omega xk_y\right) \tilde{u}^{\prime} \right] \mathrm{e}^{\mathrm{i} \mathbf{k}(t) \cdot \mathbf{r}} \right\} } \nonumber \\
  = & \Re \left\{ \left[ \partial_t \tilde{u}^{\prime} + \mathrm{i}\left( q \Omega x k_y - q \Omega x k_y \right) \tilde{u}^{\prime} \right] \mathrm{e}^{\mathrm{i} \mathbf{k}(t) \cdot \mathbf{r}} \right \} \nonumber \\
  = & \Re \left\{ \partial_t \tilde{u}^{\prime} \mathrm{e}^{\mathrm{i} \mathbf{k}(t) \cdot \mathbf{r}} \right\},
\end{align}
which cancels the aforementioned shear terms.
These solutions represent either density waves or non-wavelike perturbations of potential vorticity or entropy; there also exist axisymmetric solutions (i.e. with $k_y=0$ and $k_x = \mathrm{const}$), which are either density waves or stationary structures in potential vorticity or entropy.

\subsection{Introducing the axisymmetric shear flow} \label{sec:intro-zonal}
The structure chosen is a steady sinusoidal zonal flow accompanied by density (and therefore pressure) variations such that geostrophic balance is maintained. For that reason, it is chosen that the structure should have the form

\begin{align} \label{eq:ad-structure}
  h_{str}^\prime = & \; A_h \cos(kx) \nonumber \\
  \varv_{str}^\prime = & \; A_\varv \sin(kx) \\
  e_{str}^\prime = & \; e_0 A_e \cos(kx) \nonumber,
\end{align}
where $A_\varv$, $A_h$ and $A_e$ are the amplitudes of the imposed structure in the flow, $k$ is the structure's wavenumber and we work to first order in the amplitudes.

The expression for the axisymmetric structure in $h$ can be used to derive the form of the surface density perturbation $\Sigma^\prime$, by means of a Taylor series, making use of the relation $\Sigma_0 + \Sigma^\prime = \Sigma_0 \mathrm{e}^{h}$. This gives

\begin{equation}
  \frac{\Sigma^\prime}{\Sigma_0} = A_h\cos(kx) + \mathcal{O}(A_h^2), 
\end{equation}
which in turn allows us to find the following expression for the self-gravitional potential perturbation using Equation~\ref{eq:phi-dm}:

\begin{equation}
  \Phi_{d,m}^\prime \simeq - \frac{2\pi G \Sigma_0}{\sqrt{k_x^2 + k_y^2}} A_h\cos(kx). 
\end{equation}

The expression for the self-gravitational potential allows to deduce $A_\varv$ in terms of $A_h$ and $A_e$ making use of the equation of geostrophic balance
\begin{equation} \label{eq:geostrophic-balance}
  -2\Omega A_\varv = -2 \pi G \Sigma_0 A_h + \frac{\varv_s^2}{\gamma} k \left(A_e + A_h\right).
\end{equation}
It is worth noting that a non-SG, isothermal scenario (for which the adiabatic index is set to $\gamma=1$ and any internal energy variation is ignored, leading to $A_e=0$) would be described in an exact way by the equations provided, as the linearisation applied to the above self-gravitational potential and the non-linearity in the pressure term of Equation~\ref{eq:nonlin_u} would not be present.
Equation~\ref{eq:geostrophic-balance} yields
\begin{equation}
  A_\varv = \frac{\varv_s \kappa}{Q \Omega}A_h - \frac{\varv_s^2 k}{2\gamma \Omega} (A_e + A_h),
\end{equation}
where $Q$ represents Toomre's parameter, a measure of the disc's gravitational stability, and is defined as 
\begin{equation}
  Q \equiv \frac{\varv_s \kappa}{\pi G \Sigma_0},
\end{equation} 
where $\kappa$ is the epicyclic frequency given by $\kappa^2 = 2(2-q)\Omega^2$. 

The imposed axisymmetric structure in the three variables mentioned above also causes the presence of a structure in the potential vorticity and specific entropy, whose perturbation amplitudes can be re-written in terms of the structure's amplitude in the other quantities as follows


\begin{align}
  s^\prime = & \; \frac{P^\prime}{\gamma P} - \frac{\Sigma^\prime}{\Sigma} = \frac{1}{\gamma}\left( \frac{\Sigma^\prime}{\Sigma} + \frac{e^\prime}{e}\right) - \frac{\Sigma^\prime}{\Sigma} \nonumber \\ = & \; \frac{1}{\gamma} \left(A_e - (\gamma-1)A_h\right) \cos(kx) = A_s \cos (kx),
\end{align}
and

\begin{align} \label{eq:pot-vort-pert}
  \frac{\zeta^\prime}{\zeta} = & \; \frac{\partial_x \varv^\prime}{(2-q)\Omega} - \frac{\Sigma^\prime}{\Sigma} \nonumber \\ = & \; - \frac{1}{\kappa^2}\left[\left(A_h+A_s\right) \varv_s^2 k^2 - A_h 2\pi G \Sigma k \right] \cos(kx) - A_h \cos(kx) \nonumber \\ = & \; - \frac{1}{\kappa^2} \left[ \left( \varv_s^2 k^2 - 2\pi G \Sigma k +\kappa^2\right) A_h + \varv_s^2 k^2 A_s\right] \cos(kx) \nonumber \\ = & \; A_\zeta \cos(kx),
\end{align}
by making use of the geostrophic balance equation.
It can be noticed that the imposed structure can therefore be described through only two parameters, $A_s$ and $A_\zeta$ or, alternatively, $A_e$ and $A_h$. 

Following the introduction of the specific entropy $s$, the expression for $A_\varv$ can be futher simplified to 

\begin{equation}
  A_\varv = \frac{\varv_s \kappa}{Q\Omega} A_h - \frac{k \varv_s^2}{2\Omega} (A_s+A_h).
\end{equation}

It is important to notice that the presence of the zonal flow in the potential vorticity expression above can result in the development of an axisymmetric instability according to the Rayleigh criterion; in the shearing sheet model this occurs when the vorticity due to the zonal flow exceeds the background vorticity. Considering the un-linearised equation of the potential vorticity (Equation~\ref{eq:PV}) it is therefore possible to quickly calculate the instability criterion in the non-SG case to be $A_h \left(k \varv_s/\Omega\right)^2 > 2(2-q)=1$ for a Keplerian disc ($q=3/2$). The Rayleigh instability does however require a third dimension, and cannot therefore be triggered in our calculation.

\subsection{Introducing non-axisymmetric perturbations} \mbox{} \\
A stability analysis can be carried out by applying non-axisymmetric perturbations to the system, whose presence results in the production of a ladder of evenly spaced shearing waves as shown in Figure~\ref{fig:ladder}, which is described by

\begin{equation}  
  \label{eq:ladder-waves}
  u_i^\prime = \Re \sum_{n=-\infty}^{\infty} \left\{ \tilde{u}_n^\prime (t) \exp \left[\mathrm{i}k_{xn}(t)x + \mathrm{i}k_y y\right] \right\},
\end{equation}
and similarly for $\varv_i^{\prime}$, $h_i^\prime$ and $e_i^\prime$, where $k_{xn}(t) = k_{x0}(0) + nk + q\Omega k_y t$ is the radial wavenumber of the modes in the ladder. This gives rise to the following equations, which are solved in the analysis presented in this paper.

\begin{multline} \label{eq:ad-h}
  \mathrm{d}_t \tilde{h}_{n}^\prime = - \frac{k_y}{2} A_\varv \left(\tilde{h}_{n-1}^\prime-\tilde{h}_{n+1}^\prime \right) \\ + \frac{A_h k}{2\mathrm{i}}\left(\tilde{u}_{n-1}^\prime-\tilde{u}_{n+1}^\prime\right) - \mathrm{i} \left(k_{xn} \tilde{u}_n^\prime - k_y \tilde{\varv}_n^\prime \right) 
\end{multline}
\begin{multline} \label{eq:ad-u}
   \mathrm{d}_t \tilde{u}_n^\prime = - \frac{A_\varv k_y}{2}\left(\tilde{u}_{n-1}^\prime - \tilde{u}_{n+1}^\prime \right) + 2\Omega \tilde{\varv}_n^\prime - \mathrm{i}k_{xn} \left( \frac{\varv_s^2}{\gamma} - 2 \frac{\varv_s \kappa}{\lvert k_n\rvert Q}\right) \tilde{h}_n^\prime \\- (\gamma-1) \left[ \frac{\mathrm{i} e_0 A_e}{2} \left(k_{x,n-1} \tilde{h}_{n-1}^\prime + k_{x,n+1} \tilde{h}_{n+1}^\prime \right) \right.\\ - \left. \frac{k A_h}{2\mathrm{i}} \left(\tilde{e}_{n-1}^\prime - \tilde{e}_{n+1}^\prime \right) + \mathrm{i} k_{xn} \tilde{e}_n^\prime \right]   
\end{multline}

\begin{multline} \label{eq:ad-v}
   \mathrm{d}_t \tilde{\varv}_n^\prime = - \frac{A_\varv}{2}\left[k \left(\tilde{u}_{n-1}^\prime + \tilde{u}_{n+1}^\prime \right) + k_y \left(\tilde{\varv}_{n-1}^\prime - \tilde{\varv}_{n+1}^\prime \right) \right] \\- (2-q)\Omega \tilde{u}_n^\prime - \mathrm{i}k_{y} \left( \frac{\varv_s^2}{\gamma} - 2 \frac{\varv_s \kappa}{\lvert k_n\rvert Q}\right) \tilde{h}_n^\prime \\- (\gamma-1) \left[ \frac{\mathrm{i} e_0 A_e}{2} k_y \left(\tilde{h}_{n-1}^\prime + \tilde{h}_{n+1}^\prime \right) + \mathrm{i} k_y \tilde{e}_n^\prime \right]
\end{multline}
\begin{multline} \label{eq:ad-e}
   \mathrm{d}_t \tilde{e}_n^\prime = \frac{k e_0 A_e}{2 \mathrm{i}} \left(\tilde{u}_{n-1}^\prime - \tilde{u}_{n+1}^\prime \right) \\- \frac{A_\varv k_y}{2} \left(\tilde{e}_{n-1}^\prime - \tilde{e}_{n+1}^\prime \right) - \mathrm{i} (\gamma-1) e_0 \left\{ \left( k_{xn} \tilde{u}_n^\prime + k_y \tilde{v}_n^\prime \right) \right. \\ \left. + \frac{A_e}{2} \left[ k_{x,n-1} \tilde{u}_{n-1}^\prime + k_{x,n+1} \tilde{u}_{n+1}^\prime + k_y \left( \tilde{\varv}_{n-1}^\prime + \tilde{\varv}_{n+1}^\prime \right) \right] \right\} ,
\end{multline}
where $k_n = \sqrt{k_{xn}^2 + k_y^2}$ and $k_{x,n\pm 1} = k_{x0} + (n \pm 1)k + q\Omega k_y t$. In this description, each shearing wave is coupled to its two neighbours by the presence of the axisymmetric structure.

The above system of equations can be used for both an adiabatic and an isothermal scenario. In the latter case, it is simply necessary to set $\gamma=1$ (from $\gamma=5/3$ used in the adiabatic case) and discard any contribution arising from $e$ and $s$, along with the $e_n^\prime$ equation. This has the consequence of removing one degree of freedom from the system, with the axisymmetric structure now being described by either $A_h$ or $A_\zeta$ only.

\subsection{Method}
\label{sec:method}
As inferred from Equation~\ref{eq:ladder-waves}, the presence of an imposed axisymmetric structure on the background flow creates an infinite ladder of evenly separated shearing waves, the separation between each two neighbouring waves being the structure's wavenumber $k$. Of course, it was not possible to reproduce this infinite ladder computationally, so this had to be truncated. A brief analysis was carried out on the accuracy of the results as a function of $N$, with the total number of states considered being $2N+1$ (centred about $n=0$); this is briefly illustrated in Appendix \ref{sec:appendix}. In short, it was consistently found that a small number of states was sufficient to achieve a satisfactory accuracy and convergence in the results, and values of $2 \leq N \leq 7$ were usually found to give the best compromise of accuracy versus computational time. 

\begin{figure}
  \includegraphics[width=\columnwidth,natwidth=551,natheight=423]{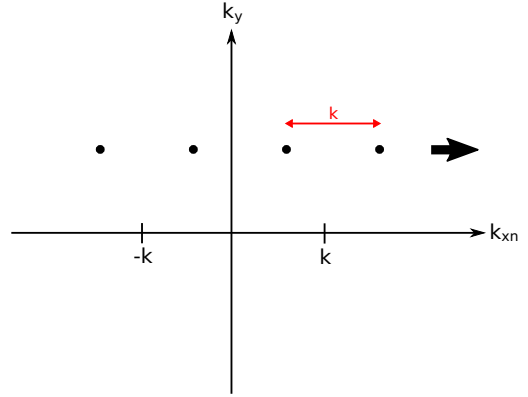}
  \caption{Representation in Fourier space of the ladder of shearing waves caused by the presence of an axisymmetric structure. These states (represented by the dots), which are evenly separated (the separation being the structure's wavenumber $k$), are sheared in the $k_{xn}$ direction matching the sign of $k_y$, as illustrated by the arrow. This means that after some time interval $\Delta t =T$, a state would have taken exactly the next place in the ladder, implying the system possesses a dynamical periodicity.}
  \label{fig:ladder}
\end{figure}

As explained previously, Equations ~\ref{eq:ad-h}--\ref{eq:ad-e} arise from the presence of an infinite ladder of evenly-spaced states (illustrated in Figure~\ref{fig:ladder} by means of dots) when an axisymmetric structure is imposed on the steady flow considered. Due to the shearing in the radial wavenumber given by $k_{xn}(t) = k_{x0}(0) + nk + q\Omega k_y t$, the states themselves shear in the positive $k_{xn}$ direction, as indicated by the arrow. This implies that there exists a time $\Delta t = T$ when the states would have moved a distance $k$ across the $k_{xn}$ axis, hence taking the exact position that the neighbour ahead of them occupied $\Delta t =T$ earlier. This means that the positions of the states of the system would look the same after every interval $\Delta t=T$, with the characteristic timescale $T=\frac{k}{q\Omega k_y}$ being called the recurrence time of the ladder of shearing waves. 
This periodicity allows us to computationally solve the steady solutions of the system in MATLAB using Floquet analysis. For this purpose, a $4(2N+1) \times 4(2N+1)$ monodromy matrix was produced by applying $4(2N+1)$ sets of initial conditions (ICs) where all variables but one are set to zero, with the non-zero variable changing in each set, to the equations of the system, evolving these for one recurrence time $T$. Each set of ICs then gives rise to a column of the matrix, with its full form obtained once all the sets of ICs have been applied.

The growth factor is determined by computing the eigenvalues of the matrix, which subsequently yields the growth rates $\lambda$ of the modes by means of 

\begin{equation}
  \lambda = \Re \left[ \frac{1}{T}\ln f \right],
\end{equation}
where $f$ is the matrix eigenvalue (i.e., Floquet multiplier) generating the largest growth rate.

The time-dependent nature of $k_{xn}$ and the finite range of modes (and hence of $k_{xn}$ values) employed meant that at the end of each recurrence time one mode exceeded the maximum value of $k_{xn}$ considered and had to be discarded; it is worth noting that this represents a form of dissipation. At the same time another mode, with negative $k_{xn}$, sheared into our wavenumber domain and had to be introduced into the matrix. This process is comparable to that used by non-linear spectral codes (eg. SNOOPY, see \citet{LesurLongaretti2005}).

\section{Stability results}
\label{sec:results}
\subsection{Isothermal case} \label{sec:iso-results}
\subsubsection{Without self-gravity}
\label{sec:no-sg}
\begin{figure}
  \includegraphics[width=\columnwidth]{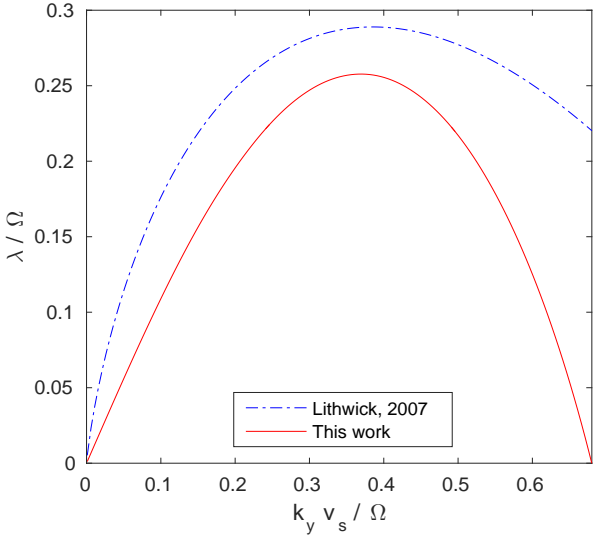}
  \caption{Comparison of the growth rate $\lambda/\Omega$ as a function of $k_y \varv_s/\Omega$ obtained in this paper (red, full line) with the results from \citet{Lithwick2007} (blue, dot-dashed), for $k \varv_s/\Omega=2$, $N=8$ and $A_h=0.25$. The poor agreement is due to the incompressibility condition, $k \varv_s/\Omega \gg1$, upon which the analysis by \citet{Lithwick2007} is based, being violated.}
  \label{fig:growthrate-k2}
\end{figure}

\begin{figure}
  \includegraphics[width=\columnwidth]{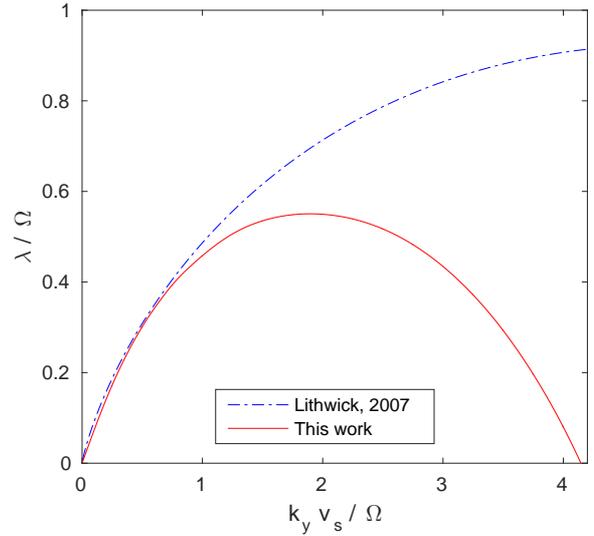}
  \caption{Similar comparison between the results obtained in this paper (red, full line) and in \citet{Lithwick2007} (blue, dot-dashed) for the growth rate $\lambda/\Omega$ as a function of the azimuthal wavenumber $k_y \varv_s/\Omega$, this time for $k \varv_s/\Omega=8$, $N=7$ and $A_h=0.05$. On this occasion both incompressibility and $k_y/k\ll1$ conditions are respected for small $k_y \varv_s/\Omega$ ($k_y \varv_s/\Omega \lesssim 1$) and the agreement in this region is satisfactory.}
  \label{fig:growthrate-k8-N7}
\end{figure}

Examples of growth rates obtained by means of the method outlined in Section~\ref{sec:method} are shown in Figures~\ref{fig:growthrate-k2} and ~\ref{fig:growthrate-k8-N7}.
These were compared to the results from \citet{Lithwick2007}, which analyses a simplified version of the problem tackled here. In particular, \citet{Lithwick2007} assumes the flow to be incompressible (which in this analysis corresponds to $k \varv_s/\Omega\gg1$, so that the scale of the structure is much smaller than the disc scale height) and constrains the analysis to the $k_y/k\ll1$ limit. In said limit, \citet{Lithwick2007} finds the growth rate to be described by the expression

\begin{equation}
\label{eq:lithwick-growth-rate}
  \lambda = \frac{q\Omega k_y}{k} \ln\left(\frac{\pi A_h \varv_s^2 k^3}{4q\Omega^2 k_y}\right),
\end{equation}
which was derived by converting an analogous expression from said work.

Figures~\ref{fig:growthrate-k2} and ~\ref{fig:growthrate-k8-N7} show a comparison between the growth rate as a function of $k_y \varv_s/\Omega$ between the analysis described in this paper (red, full lines) and that carried out by \citet{Lithwick2007} (blue, dashed). 

As shown in Figure~\ref{fig:growthrate-k2}, the two calculations do not appear to be consistent with each other for a moderate value of $k \varv_s/\Omega$. This is because, although the $k_y/k\ll 1$ limit is attained for very small values of $k_y \varv_s/\Omega$, the incompressibility condition $k \varv_s/\Omega \gg1$ is violated. Therefore in this case Equation~\ref{eq:lithwick-growth-rate} is not valid.

In Figure~\ref{fig:growthrate-k8-N7}, on the other hand, a higher value of $k \varv_s/\Omega$ is considered to make sure that the incompressibility assumption taken in \citet{Lithwick2007} holds and a level of agreement is reached. In fact, the agreement is excellent for $k_y \varv_s/\Omega \lesssim 1$; as $k_y \varv_s/\Omega$ increases further, however, the $k_y/k\ll1$ limit begins to be violated and the two curves diverge, as expected. Due to the satisfactory degree of agreement between the two analyses in the region of interest, it is safe to state that the instability detected is a Kelvin-Helmholtz instability -- exactly like that found by \citet{Lithwick2007}. 

The issue of convergence as a function of the total number of states $2N+1$ is explored in the Appendices, with Appendix ~\ref{sec:app-negative} analysing the effects on neutral/decaying modes and Appendix ~\ref{sec:app-res} looking at the convergence of the growth rate themselves. However all results plotted here are converged.

\paragraph*{Stability analysis as a function of the structure's properties}
\label{sec:A_k_contour}
In order to better understand how the presence of an axisymmetric structure affects the stability of the flow, an analysis of the growth rates produced was carried out for a range of values for the structure's dimensionless wavenumber, $k \varv_s/\Omega$, and amplitude, $A_h$. The growth rate values were maximised with respect to $k_y$ by calculating them at a series of $\sim20$ evenly spaced $k_y \varv_s/\Omega$ values in the interval $k_y \varv_s/\Omega \in (0,k\varv_s/\Omega]$.

\begin{figure}
  \includegraphics[width=\columnwidth]{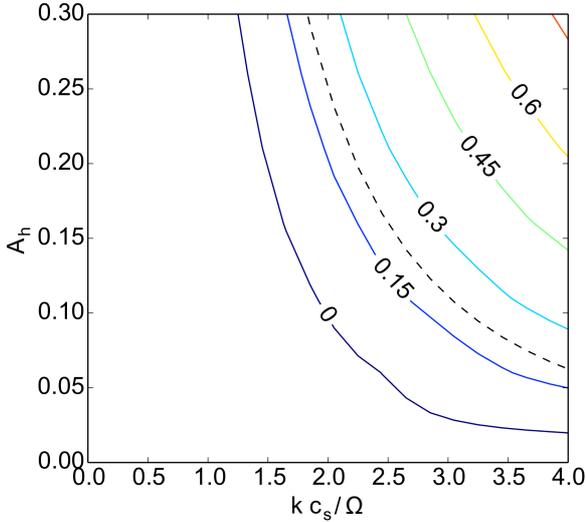}
  \caption{Contours of the growth rate (in units of $\Omega$) maximised with respect to $k_y$ in the $k \varv_s/\Omega$ -- $A_h$ plane with no self-gravity. The strength of the growth increases both with increasing $k \varv_s/\Omega$ and with increasing $A_h$; the mere presence of the axisymmetric structure is however not always sufficient to trigger an instability; indeed, even for an amplitude of $A_h=0.3$, no instability is detected if the wavelength of the structure is sufficiently long. The dashed line represents the set of parameters for which $A_h (k\varv_s/\Omega)^2=1$, with the flow being unstable to axisymmetric disturbances according to the Rayleigh criterion on the right of the curve.}
  \label{fig:A_k_contour}
\end{figure}

Figure~\ref{fig:A_k_contour} shows the growth rate contours in the $k \varv_s/\Omega$ -- $A_h$ plane; the strength of the instability is found to increase with both parameters. However, although a small amplitude is sufficient to trigger an instability for values of $k \varv_s/\Omega$ exceeding $\sim 3.5$, a rather large amplitude doesn't appear to be enough to make the system unstable if the wavelength of the structure is sufficiently long. 
This seems to imply that although a minimum would be present in the potential vorticity profile (\citet{LinPapaloizou2011, LovelaceHohlfeld2013}), its presence alone is not enough to imply the flow is unstable. Instead, a constraint on the shape of the minimum must be applied to find out whether it triggers an instability. This ties in with the work by \citet{Lovelaceetal1999}, where the height-to-width ratio of a minimum in the potential vorticity $\zeta$ must exceed a critical value for an instability to occur. 
Figure~\ref{fig:A_k_contour} also presents a dashed curve which indicates the set of points that satisfy the condition $A_h (k\varv_s/\Omega)^2=1$; as seen in Section~\ref{sec:intro-zonal}, this represents the marginal stability condition given by the Rayleigh criterion. Although this type of instability requires a third dimension to operate, it is important to be aware of its presence and its operating region as it can limit the properties of the zonal flow considered.  What is clear is that the non-axisymmetric KH instability can operate in a Rayleigh stable regime (i.e., on the left of the dashed curve), meaning the length-scale and amplitude of the zonal flow are constrained by the non-axisymmetric instability. 

The behaviour of the zero growth rate contour (and others) at larger $k \varv_s/\Omega$ values can be inferred through a comparison with the incompressible work of \citet{Lithwick2007}, which states that the growth rate of the non-axisymmetric KH instability is given by Equation~\ref{eq:lithwick-growth-rate}. The value of $A_h$ can then be extrapolated from the equation at a given $k \varv_s/\Omega$ for the corresponding azimuthal wavenumber value. 

\begin{figure}
  \includegraphics[width=\columnwidth]{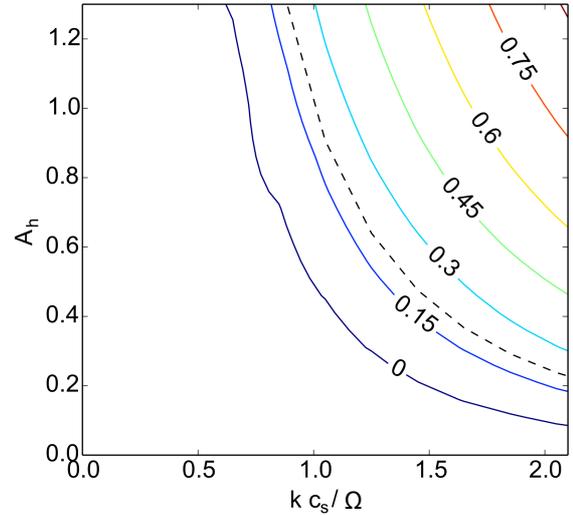}
  \caption{Further analysis on the growth rate plot of Figure \ref{fig:A_k_contour} as a function of $k \varv_s/\Omega$ and $A_h$ for the $k \varv_s/\Omega\lesssim2$ range. Even for this increased range in $A_h$ the system remains stable if the wavelength of the applied axisymmetric structure is sufficiently long. Moreover, the larger $A_h$ range is still not sufficient to determine whether the instability threshold would retain a weak $A_h$--dependence or if it would eventually exhibit an asymptotic behaviour. The dashed line again represents the marginal stability line (where $A_h (k\varv_s/\Omega)^2=1$) given by the Rayleigh criterion.}
  \label{fig:A_k_contour_2}
\end{figure}

Further analysis was carried out to focus on the low--$k \varv_s/\Omega$ part of the plane, as shown in Figure~\ref{fig:A_k_contour_2}. This is to investigate whether the $k\varv_s/\Omega$ threshold value -- below which no instability can be triggered -- eventually exhibits an asymptotic behaviour and becomes independent of $A_h$. It was possible to achieve large values of $A_h$ thanks to the imposed axisymmetric structure being described in an exact way in the isothermal, non-self-gravitating case. Regardless of the rather large values of $A_h$ investigated, it doesn't appear possible to tell whether the threshold $k \varv_s/\Omega$ value would continue to indicate a weak dependence on the amplitude $A_h$ or become an asymptote. However by looking at same parameter range as Figure~\ref{fig:A_k_contour} with different values of $\gamma$, it was possible to establish that the threshold value for the zonal flow stability is independent of the adiabatic index value. This threshold can be interpreted as a critical wavelength of zonal flow, with structures possessing a wavelength $\gtrsim 8H$ (where $H=\varv_s/\Omega$ is the scale height of the disc) being stable.

\subsubsection{With self-gravity}
The consideration of the effects of self-gravity upon a system on which an axisymmetric structure has been imposed represents a clear extension of the work carried out by \citet{Lithwick2007}, and its addition could result, under the right conditions, into the detection of a different type of instability. A related stability analysis has been carried out by \citet{LovelaceHohlfeld2013}, who observed instabilities, both in the presence and in the absence of self-gravity, when a simpler, single structure with a discontinuous profile is present in the density.

\begin{figure}
	\includegraphics[width=\columnwidth]{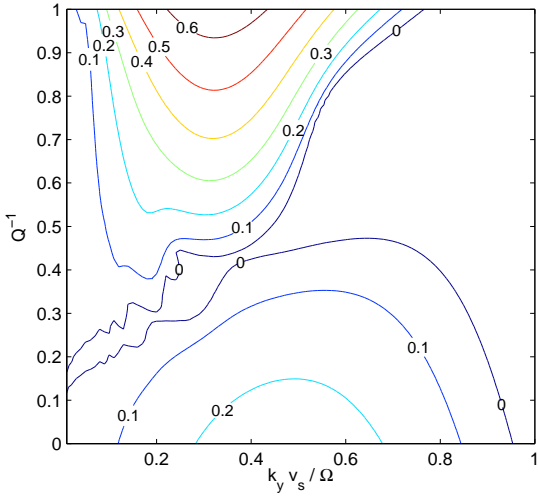}
    \caption{Growth rate contours in the $k_y \varv_s/\Omega$ -- $Q^{-1}$ plane for $k \varv_s/\Omega = 3$ and $A_h = 0.1$. The graph clearly shows two distinct types of instability: a KH instability taking place at low $Q^{-1}$, while a gravitationally-induced instability kicks in at intermediate values of $Q^{-1}$, following a small area of marginal stability.}
    \label{fig:iso-growth-rate-contour}
\end{figure}

In order to be able to discern the presence of an additional type of instability, alongside the KH instability discussed in Section~\ref{sec:no-sg}, the growth rates were calculated for various self-gravity strengths. This allowed the construction of a contour plot for the growth rate as a function of both $k_y \varv_s/\Omega$ and $Q^{-1}$, which is presented in Figure~\ref{fig:iso-growth-rate-contour}.

For $Q^{-1}=0$ the KH instability discussed in Section~\ref{sec:no-sg} is again found here, initially centred around $k_y \varv_s/\Omega=0.5$. As self-gravity is introduced, the strength of the KH instability gradually wanes while its efficacy seems to shift to smaller scales ($0.6 \lesssim k_y \varv_s/\Omega \lesssim 0.7$). Upon increasing the strength of self-gravity a little further, a region of marginal stability is achieved, with the $Q^{-1}$ value necessary for stability strongly varying as a function of the azimuthal wavenumber. As $Q^{-1}$ is further increased, a non-axisymmetric instability of gravitational nature develops, eventually settling at a preferred wavenumber of $k_y \varv_s/\Omega\simeq 0.35$. For $Q^{-1} > 1$ (ie, at the top of the shown range) SG causes the system to undergo an axisymmetric instability, even in the absence of an imposed structure.

\begin{figure}
	\includegraphics[width=\columnwidth]{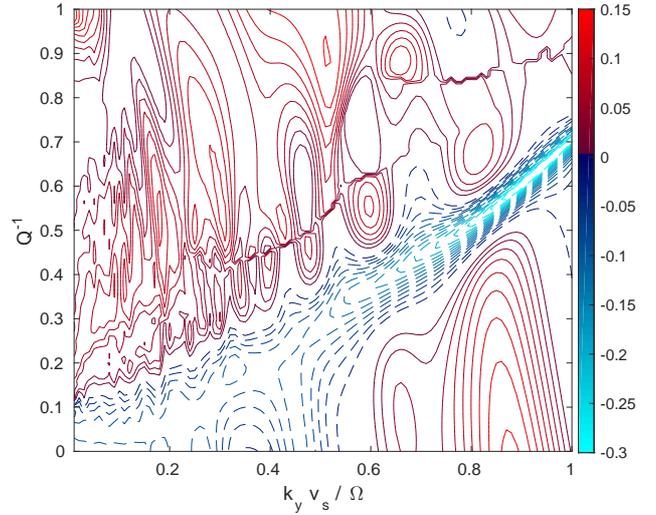}
  \caption{Normalised time-averaged correlation integral $I$ of the wave energy with the potential vorticity $\zeta$ in the $k_y \varv_s/\Omega$ -- $Q^{-1}$ plane for $k \varv_s/\Omega = 3$ and $A_h = 0.1$. The majority of positive (red, full contours) and negative (blue, dashed) values in the top and bottom halves of the plot, respectively, confirms the presence of two distinct types of instability. It also illustrates how the gravitationally-induced instability is associated with a maximum in the potential vorticity, while the KH instability is localised around a minimum.}
  \label{fig:I}
\end{figure}

\paragraph*{Analysing the nature of the gravitationally-induced instability}
A plethora of dynamical instabilities are possible in accretion discs, some of which are associated with extrema in potential vorticity, or alternatively steep gradients in the surface density \citep{PapaloizouPringle1985, PapaloizouPringle1987, PapaloizouLin1989, PapaloizouSavonije1991, Lovelaceetal1999, LinPapaloizou2011, LovelaceHohlfeld2013}.
In order to establish whether the gravitationally-induced instability was indeed caused by extrema in the potential vorticity $\zeta$, the correlation integral $I$ of the potential vorticity due to the axisymmetric structure with the wave's energy was analysed. The acoustic energy is given by

\begin{equation}
  E = \frac{\Sigma}{2}\left(u^{\prime \, 2} + \varv^{\prime \, 2}\right) + \frac{P^{\prime \, 2}}{2\gamma P}
\end{equation}
which in isothermal conditions simplifies to
\begin{equation}
  E \propto u^{\prime \, 2} + \varv^{\prime \, 2} + \varv_s^2 h^{\prime \, 2}.
\end{equation}
Considering the correlation integral between the azimuthally averaged acoustic energy and Equation~\ref{eq:pot-vort-pert}, we obtain

\begin{equation}
  I \propto - \Re \left\{ \sum_{n} \tilde{u}^\prime_n \tilde{u}_{n+1}^{\prime \, *} + \tilde{\varv}_n^\prime \tilde{\varv}_{n+1}^{\prime \, *} + \varv_s^2 \tilde{h}^\prime_n \tilde{h}_{n+1}^{\prime \, *} \right\}
\end{equation}
where $\Re$ indicates that we're only interested in the real part of the expression enclosed within curly brackets and the minus sign arises simply from the convenience of having positive values of $I$ associated with potential vorticity maxima, rather than minima. Its time-average was plotted as a function of both $Q^{-1}$ and $k_y \varv_s/\Omega$, as shown in Figure~\ref{fig:I}.

\begin{figure*}
  \centering
  \begin{minipage}{.99\textwidth}    
    \subfloat[]{\includegraphics[width=.47\columnwidth]{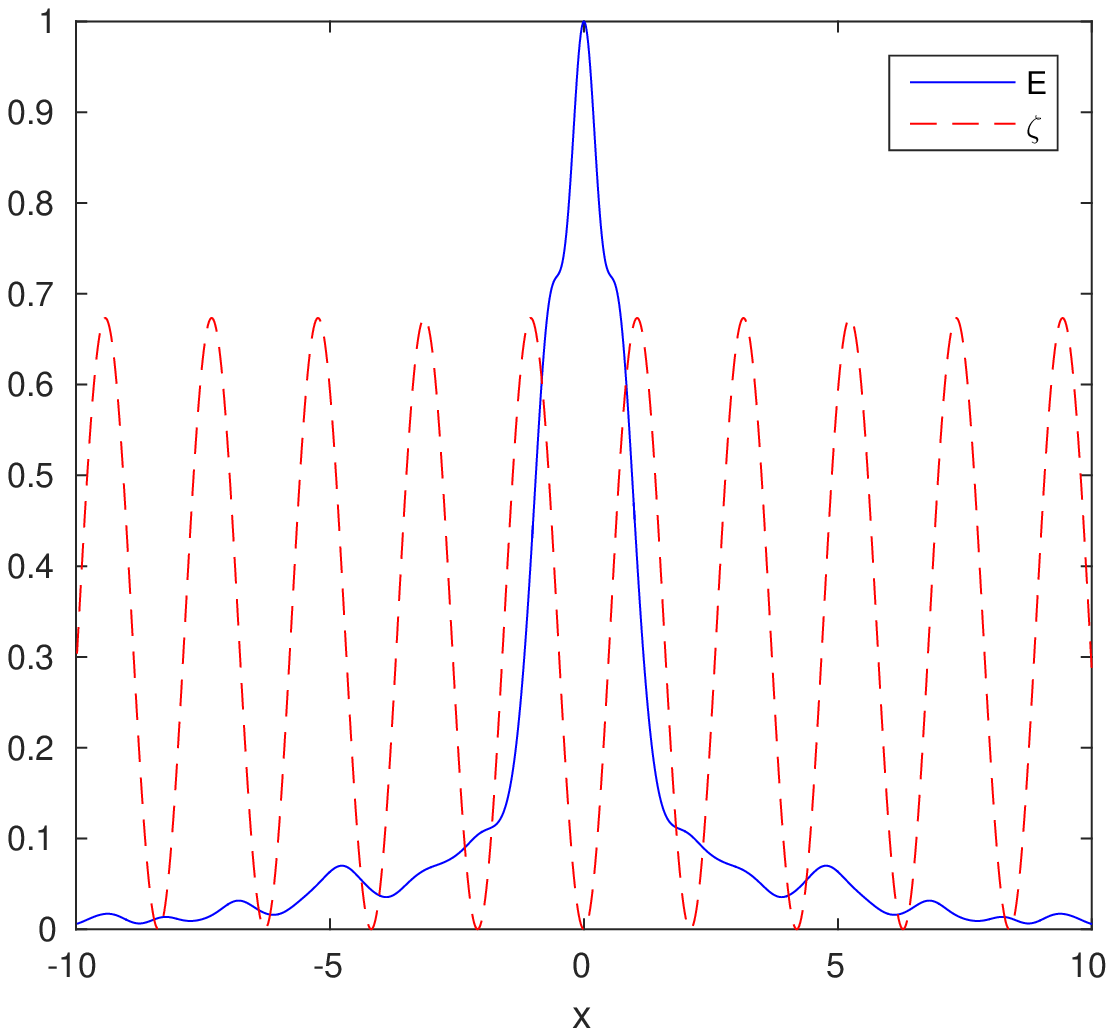}}
    \hspace*{.05\columnwidth}
    \subfloat[]{\includegraphics[width=.47\columnwidth]{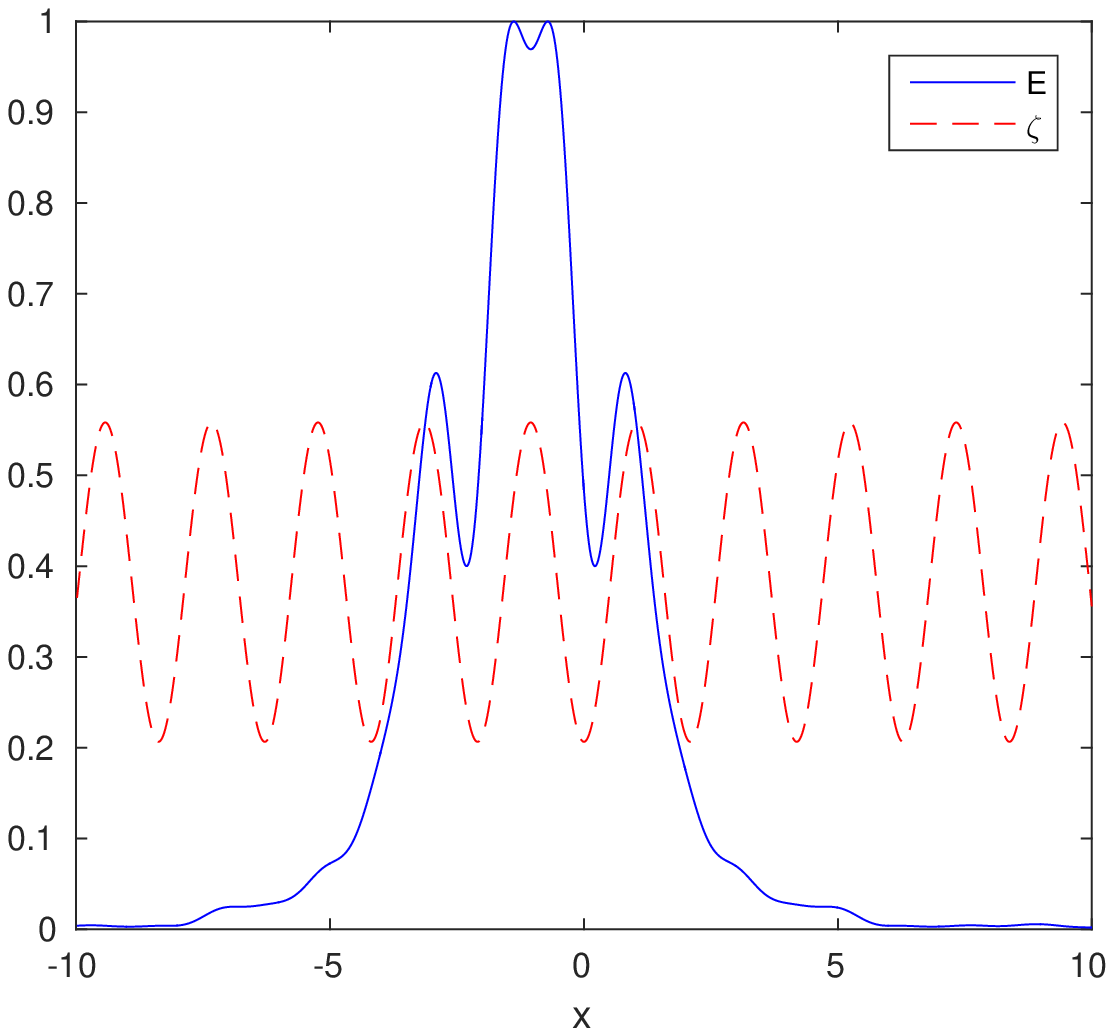}}
    \caption{Real space visualisation of the normalised energy (full, blue line) associated with \textbf{\textit{(a)}} the Kelvin-Helmholtz instability ($k_y \varv_s/\Omega=0.4$, $Q^{-1}=0$) and \textbf{\textit{(b)}} the gravitational instability ($k_y \varv_s/\Omega=0.3$, $Q^{-1}=0.9$) modes for $k\varv_s/\Omega=3$, $A_h=0.1$. A scaled up version of the potential vorticity $\zeta$ is also shown (dashed, red line) as a means of confirmation for the modes being localised around a minimum and a maximum of the potential vorticity, respectively. The degree of localisation is believed to be intrinsically linked to the properties of the induced axisymmetric structure. }
    \label{fig:real_space_energy_PV}
  \end{minipage}
\end{figure*}

The most immediately visible feature of Figure~\ref{fig:I} is the presence of two distinct zones in the $k_y \varv_s/\Omega$ -- $Q^{-1}$ plane, roughly separated by a diagonal line, which show a majority of positive and negative $I$ values, respectively. This therefore corroborates the implication inferred from Figure~\ref{fig:iso-growth-rate-contour} as to there being two separate modes of instability.
The mostly positive (red, full contours) contour values in the top half of the plot suggest that the gravitationally-induced instability is centred on a maximum in the potential vorticity, meaning the negative (blue, dashed) values linked to the KH instability imply it is associated with a minimum in $\zeta$, instead. There also appears to be a sizeable fraction of the KH instability region linked with a maximum in the PV, whose presence is explained below.
Combining the information gathered from Figures~\ref{fig:iso-growth-rate-contour} and \ref{fig:I}, it is therefore possible to deduce that the introduction of self-gravity stabilises the instability (KH instability) associated with a minimum in the potential vorticity until the system is brought to (marginal) stability, while at the same time triggering a new type of instability, this time associated with a maximum in $\zeta$. This is in broad agreement with the results obtained by \citet{LinPapaloizou2011, LovelaceHohlfeld2013}.

These results are confirmed by the visualisation of the modes belonging to each instability in real space, which was obtained by reconstructing a mode's eigenfunction using Fourier transforms making use of the periodicity of the shearing wave ladder and the Poisson summation formula, as shown in Figure~\ref{fig:real_space_energy_PV}. This shows the energy (full, blue lines) of a mode associated with \textbf{\textit{(a)}} the KH instability ($k_y \varv_s/\Omega=0.4$, $Q^{-1}=0$) and \textbf{\textit{(b)}} the gravitational instability ($k_y \varv_s/\Omega=0.3$, $Q^{-1}=0.9$) and, as a reference, a scaled up version of the potential vorticity $\zeta$ (dashed, red lines). The energy of the modes is well localised near an extremum of the potential vorticity, with the KH instability and gravitational instability modes centred around a minimum and a maximum in $\zeta$, respectively. Both instability modes are centred around the corotation radius ($x_{\mathrm{CR}}=0$ and $x_{\mathrm{CR}}=-\pi/3$ for KH and gravitational instabilities), with density waves seen propagating away from corotation beyond the Inner and Outer Lindblad Resonance radii. This mode visualisation technique also allows us to analyse the meaning of positive $I$ values (i.e. linked to a PV maximum) in the KH part of the plane of Figure~\ref{fig:I}. The modes corresponding to this area of $I>0$ are still found to be centred around a PV minimum, however the mode's energy is much less localised around the extremum as well as the secondary peaks due to the propagation of density waves being much more prominent; both of these effects contribute to the total value of $I$ being positive. 

We also looked into the possibility of each instability being potentially identified by the amounts of compressibility and vorticity present in the part of the $k_y \varv_s/\Omega$--$Q^{-1}$ plane it occupies; it might be expected that the gravitationally-induced and KH instabilities should have a predominantly compressible and vortical character, respectively. For this reason, we looked at the function
\begin{equation}
  \Upsilon  = \ln \left( \frac{\int \lvert \nabla \cdot \boldsymbol\varv \rvert^{2} dA}{\int \lvert \nabla \times \boldsymbol\varv \rvert^{2} dA}\right),
\end{equation}
which in Fourier space, according to Parseval's theorem, is transformed to
\begin{equation}
  \Upsilon  = \ln \left( \frac{\sum\limits_n \lvert k_{xn} u^\prime_n + k_y \varv^\prime_n\rvert^2}{\sum\limits_n \lvert k_{xn} \varv^\prime_n - k_y u^\prime_n\rvert^2}\right),
\end{equation}
where $n \in \left[-N,N\right]$.

\begin{figure}
	\includegraphics[width=\columnwidth]{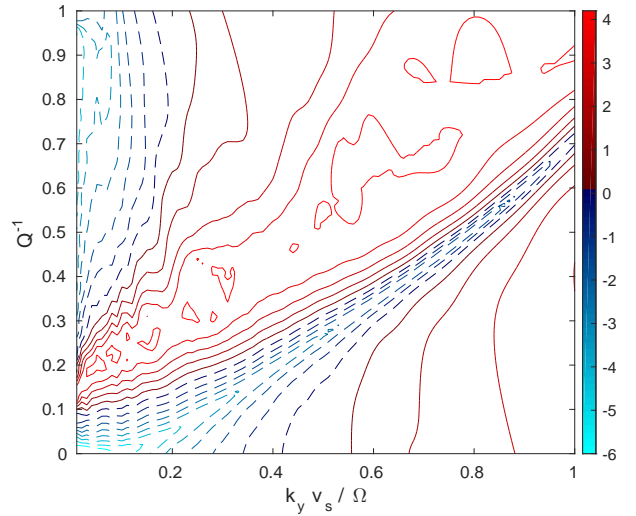}
  \caption{Contour plot of the quantity $\Upsilon$ for $k\varv_s/\Omega=3$ and $A_h=0.1$. Two somewhat distinct regions are present, with the half where the gravitational instability sits showing a compressive behaviour (red, full contours) while the lower half is dominated by vortical motions (blue, dashed contours).}
  \label{fig:log_compr_to_vort}
\end{figure}

Figure~\ref{fig:log_compr_to_vort} shows the value and sign of $\Upsilon$ in the domain considered. It is again possible to notice how the plane is roughly split into two areas by a diagonal line stretching from $\sim(k_y\varv_s/\Omega=0, Q^{-1}=0.1)$ to $(1,0.7)$, with the lower region (which is where the KH instability sits) showing a significant amount of vorticity (blue, dashed contours) and the upper region (where the gravitationally induced instability lies) dominated by compressive motions (red, full contours), as expected. Both instabilities however appear to have a vortical nature in the low $k_y \varv_s/\Omega$ region, underlying the importance of vorticity (and potential vorticity) in the problem; compressibility, on the other hand, becomes increasingly important as the azimuthal wavenumer tends to $k_y\varv_s/\Omega \sim 1$, where the density wave production by vortical motions is maximised \citep{HeinemannPapaloizou2009}.

\subsection{Adiabatic case}
\begin{figure}
	\includegraphics[width=\columnwidth]{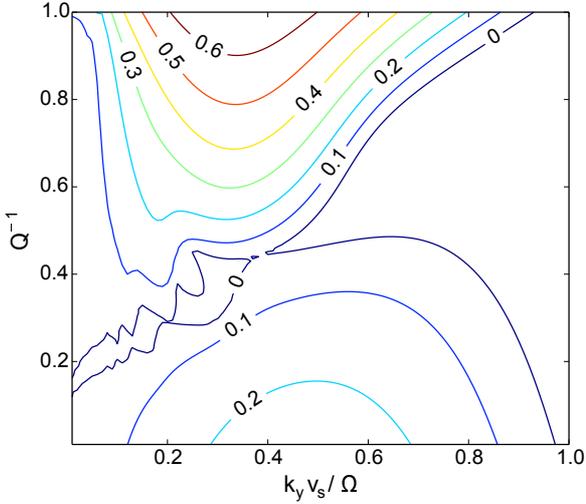}
  \caption{Growth rate contour analysis similar to Figure~\ref{fig:iso-growth-rate-contour} but in adiabatic conditions, with $\gamma=5/3$, and assuming $A_s=0$. In order to make a direct comparison with Figure~\ref{fig:iso-growth-rate-contour}, $A_h=0.1$ is assumed throughout the plane. The plot looks similar to the isothermal case, with the KH instability in particular being particularly insensitive to the different $\gamma$ value; the gravitational instability also looks similar to that observed in Figure~\ref{fig:iso-growth-rate-contour}, with the exception that it is active for a broader range of wavelengths.}
  \label{fig:ad-As0}
\end{figure}

An important difference between the adiabatic and isothermal regimes is that in the former case a structure can also be present in the entropy, as well as in the potential vorticity, which could provide key differences in the dynamics and stability of the flow. As a means of confirmation of previous results, calculations were made in the adiabatic scenario by removing any structure in the entropy (i.e., setting $A_s=0$), and the results obtained were very similar in nature to those found in the isothermal scenario for the corresponding set of parameters, as shown in Figure~\ref{fig:ad-As0}. A comparison with Figure~\ref{fig:iso-growth-rate-contour} reveals the KH instability being insensitive to the different $\gamma$ value while small differences can be spotted for the gravitational instability, which appears to be active for a more extended range of wavelengths compared to the isothermal case, reaching larger $k_y \varv_s/\Omega$ values. 

The work by \citet{Lovelaceetal1999} analysed the onset of instabilities in thin, non-self-gravitating Keplerian accretion discs and detected a linear Rossby wave instability when a function that is a combination of entropy and potential vorticity presents a local maximum in its radial profile. More interestingly, and perhaps surprisingly, they also found a Rossby wave instability when the local maximum previously mentioned is only present in the entropy profile; according to their analysis, waves become trapped in the vicinity of the entropy maximum and provide outward angular momentum transport.

\begin{figure}
	\includegraphics[width=\columnwidth]{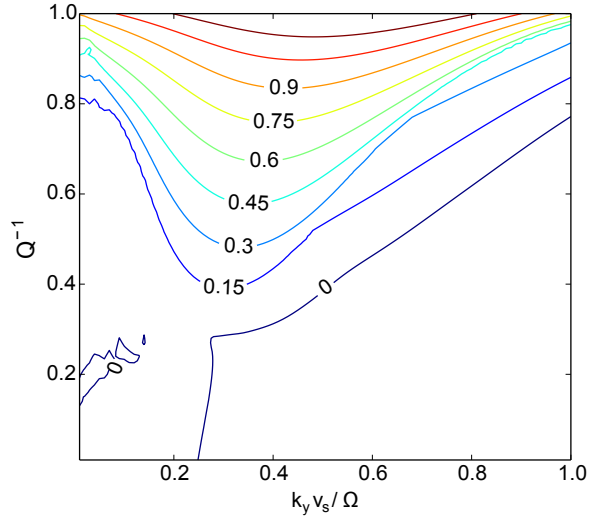}
  \caption{Growth rate contours for $k \varv_s/\Omega=2$, $\gamma=5/3$, $A_{\zeta}=0$ and $A_s=0.25$ showing the gravitationally-induced instability having spread to low values of $Q^{-1}$ and an instability being present for $Q^{-1}=0$, too.}
  \label{fig:ad-k5}
\end{figure}

Here a similar approach was taken by setting the perturbation amplitude for the axisymmetric structure in the entropy and potential vorticity, although the entropy here is defined in a slightly different manner compared to \citet{Lovelaceetal1999}. To verify the claims of \citet{Lovelaceetal1999}, the structure in the potential vorticity $\zeta$ was replaced with one in the entropy $s$, which in turn meant that the Kelvin-Helmholtz instability seen in Section~\ref{sec:iso-results} was no longer active, leaving only the instability induced by self-gravity. The value of $A_s$ was then gradually increased, resulting in the gravitational instability operating at increasingly smaller values of $Q^{-1}$; as $A_s$ was increased further, an instability for $Q^{-1}=0$ also developed, as shown in Figure~\ref{fig:ad-k5}. As it can be seen, this does not present two distinct instabilities like the equivalent plot in the isothermal case; it is therefore unclear whether the $Q^{-1}=0$ and $Q^{-1}\neq0$ instabilities are of the same type, or whether they are of different nature but appear uniformly merged. 

\begin{figure}
	\includegraphics[width=\columnwidth]{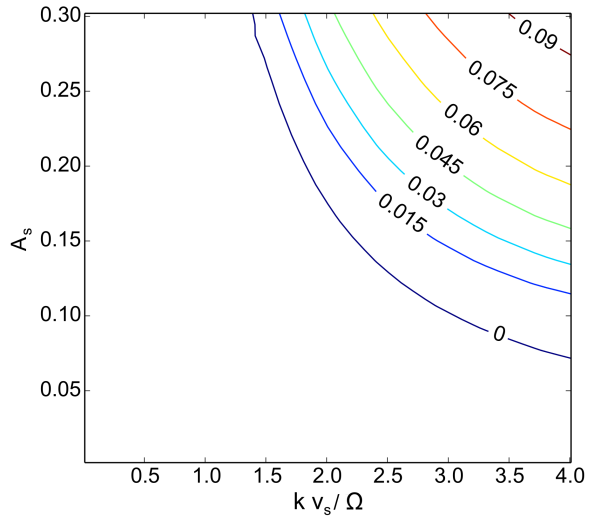}
  \caption{Contours of the growth rate (in units of $\Omega$) maximised with respect to $k_y$ in the $k \varv_s/\Omega$ -- $A_s$ plane for $A_\zeta=0$, for non-SG conditions. As in the isothermal case again the growth rate strength increases with increasing $k \varv_s/\Omega$ and $A_s$, although no instability is detected for $A_s \lesssim 0.075$ or $k \varv_s/\Omega \lesssim 1.5$ in the chosen range. In this case the growth rates achieved are roughly 1 order of magnitude smaller than in the isothermal case. The flow is in this case Rayleigh stable for all parameter values.}
  \label{fig:ad-Ask_contour}
\end{figure}

A similar analysis as that in Section~\ref{sec:A_k_contour} was carried out for the adiabatic case, again neglecting self-gravity, to investigate how the instability detected in Figure~\ref{fig:ad-k5} depends on the structure's wavenumber and the amplitude $A_s$; Figure~\ref{fig:ad-Ask_contour} shows the resulting plot for the case where $A_\zeta=0$.
Although the growth rate behaves in a qualitatively similar way to that plotted in Figure~\ref{fig:A_k_contour}, in that the growth rates, where present, increase with both increasing $k \varv_s/\Omega$ and $A_h$, there are some differences between the two plots. Firstly, an instability is slightly more difficult to trigger in the absence of a PV structure, for no instability is detected for $A_s \lesssim 0.075$ or $k \varv_s/\Omega\lesssim 1.5$ in the chosen range. Secondly, when an instability is triggered it appears to be somewhat weaker, with the growth rates detected being roughly 1 order of magnitude smaller than those found in the isothermal scenario. It is worth noting that the system is stable according to the Rayleigh criterion for all values of the structure's wavenumber and amplitude, as a uniform potential vorticity (i.e. $A_\zeta=0$) means the vorticity is always positive.

\begin{figure}
	\includegraphics[width=\columnwidth]{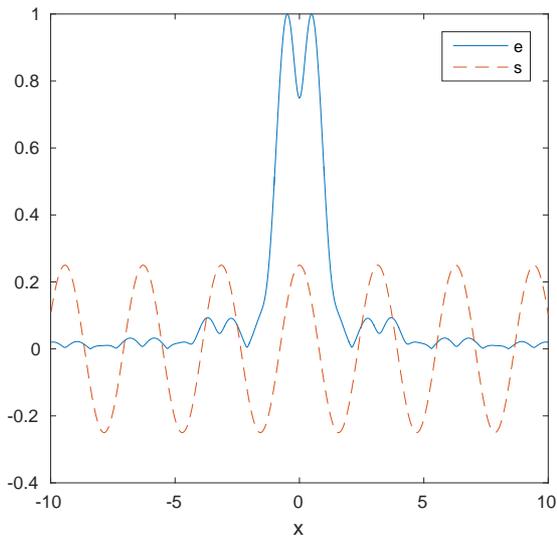}
  \caption{Real space visualisation of the normalised internal energy $e$ (full, blue line) associated with the instability detected for non-SG conditions, for $k_y \varv_s/\Omega=0.075$, $k \varv_s/\Omega=2$, $A_s=0.25$, $A_\zeta=0$ and $\gamma=5/3$. The entropy perturbation is also shown (dashed, red line) to show that this type of instability is localised around a maximum in $s$.}
  \label{fig:real_space_energy_entropy}
\end{figure}

The Poisson summation formula method was again employed to obtain the real space visualisation of a mode associated with the instability observed for non-SG (from Figure~\ref{fig:ad-k5}), which is shown in Figure~\ref{fig:real_space_energy_entropy}. This shows the mode's internal energy $e$ (full, blue line) being well localised near a maximum in the specific entropy $s$ (dashed, red line) for $k_y\varv_s/\Omega=0.075$, $k \varv_s/\Omega=2$, $A_s=0.25$ and $A_\zeta=0$. 
Although the smaller growth rates produced by this instability imply that it would only play a minor role in the dynamics of the system, we feel it is something worth pursuing.
It would also be intriguing to extend the $k \varv_s/\Omega$ range in Figure~\ref{fig:ad-Ask_contour} to higher values to check whether the threshold $A_s$ value eventually becomes independent of the wavelength; however, this would result in considering wavelengths smaller than the scale height of the disc, which is probably an inadequate regime for the current 2D model.

\section{Conclusions}
\label{sec:conclusions}
We carried out a linear stability calculation of a steady Keplerian flow on which an axisymmetric structure has been imposed both in the absence and presence of self-gravity. For the isothermal case with no self-gravity, the growth rates of the non-axisymmetric instability induced by the presence of the axisymmetric structure in the potential vorticity $\zeta$ were compared to the incompressible calculation carried out by \citet{Lithwick2007}. The growth rates from the two calculations agree well for structures and disturbances of short wavelength, although a discrepancy appears as the azimuthal wavenumber increases due to a simplification in the model employed by \citet{Lithwick2007}. Thanks to the level of agreement reached, the non-axisymmetric instability found was pinpointed to be of the Kelvin-Helmholtz type.

The strength of the Kelvin-Helmholtz (KH) instability was investigated as a function of the structure's amplitude and its wavelength. Zonal flows were found to be stable to non-axisymmetric instabilities, even for large amplitudes, as long as their wavelength $\gtrsim 8H$, with $H$ being the disc scale height; the result was found to be independent of the adiabatic index $\gamma$. Zonal flows having wavelength smaller than that critical value can undergo a KH instability, which would break the structure up into vortices. On the other hand, long-lived zonal flows observed in magnetorotational instability (MRI) simulations, can be explained either by the structure possessing a sufficiently long (and hence stable) wavelength, or by the box employed being too small in the azimuthal direction, preventing the capture of the instability. In the latter case, the results presented in this paper suggest that the aspect ratio of the box should be $L_y/L_x \gtrsim 2-2.5$ in order to successfully capture the KH instability. 
On the other hand, if the zonal flow wavelength is less than $\sim2H$, a small zonal flow amplitude is enough to trigger a KH instability. The Rayleigh criterion was also looked at as a potential way of constraining the size of the zonal flow, although its resulting axisymmetric instability needs a 3D model to operate. It was found that it is possible for the KH instability to operate in a Rayleigh stable regime, meaning that the length-scale and the amplitude of any zonal flow would be limited by the non-axisymmetric instability. As highlighted by \citet{Lithwick2007, Lithwick2009} the KH instability observed leads to the formation of vortices in the disc, while no direct widening of the zonal flow is seen; it is however possible that the non-linear dynamics of the vortices induced by the instability might indirectly cause the formation of a wider, stable zonal flow configuration.

Self-gravity was subsequently introduced and a second, distinct type of non-axisymmetric instability was detected. This newly pinpointed instability was of a gravitational nature and was found to be linked -- by considering the correlation integral of the wave's energy with the potential vorticity -- to maxima in the potential vorticity, while the previously detected KH instability is associated with minima in $\zeta$; this agrees with results in the literature, among which are \citet{LinPapaloizou2011, LovelaceHohlfeld2013}. This was confirmed by looking at the $x$--profile of the energy for modes associated with each instability. A brief analysis on the amount of compressibility and vorticity associated with each instability was also carried out, and this in general confirmed the initial expectation that the KH and gravitational instabilities should present an excess of vorticity and compressibility, respectively. However in the long $2\pi/k_y$ regime, both instabilities showed a predominance of vorticity, underlining the importance of the potential vorticity's role in this problem.

An ideal adiabatic case was also considered to analyse the claim by \citet{Lovelaceetal1999} that a non-self-gravitating linear Rossby wave instability can be triggered by the presence of a local maximum in the entropy profile but not in the potential vorticity, which is usually a requirement for the development of a Rossby wave instability. A gravitational non-axisymmetric instability was detected, and its activation point shifted down to weaker gravity conditions as the amplitude of the entropy structure was increased. Eventually an instability was detected for no self gravity, therefore confirming the presence of such an instability as claimed by \citet{Lovelaceetal1999}. It was however unclear whether this instability, which was found to be localised near an entropy maximum using the Poisson summation formula method, was of a different type from the one obtained for non-zero self-gravity. A study into the exact nature of this instability was beyond the scope of this paper, but an analysis of its strength as a function of the zonal flow's amplitude and wavelength was carried out; the result was qualitatively similar to the KH instability in the isothermal case, except for the entropy-induced instability being weaker by an order of magnitude. It could also be triggered for a smaller range of amplitude and wavenumber values. In this case the flow is stable to axisymmetric disturbances according to the Rayleigh criterion for all values of the structure's amplitude and wavelength, so once again the properties of the zonal flow are limited by the non-axisymmetric instability.  

This work is based on an ideal model of the problem at hand, where the imposed zonal flow is assumed to be in a steady state with diffusive and -- in the adiabatic case -- cooling effects being neglected. While this ideal situation allows us to consider and analyse the dynamics of the flow without having to untangle them from secondary effects, it obviously represents an unrealistc scenario in the case of a real life accretion disc. For that reason, the next step is to expand this work by including effects such as those mentioned above and analysing their effects upon the evolution and stability of the system. 

\section*{Acknowledgements}
We would like to thank the reviewer, Andrew Youdin, for providing a detailed and constructive set of comments.
The research was carried out thanks to the funding provided by the Science \& Technology Facilities Council (STFC).




\bibliographystyle{mnras}
\bibliography{linear-analysis-paper} 




\appendix

\section{Effects of resolution} \label{sec:appendix}
As discussed in Section~\ref{sec:method}, the presence of an imposed axisymmetric structure upon the flow of the Keplerian disc creates an infite ladder of evenly-separated shearing waves, which is of course impossible to represent within a computer simulation. For that purpose, a brief analysis on how the number of states considered affects the results was carried out.

\subsection{Energy conservation}
\label{sec:app-negative}
\begin{figure}
	\includegraphics[width=\columnwidth]{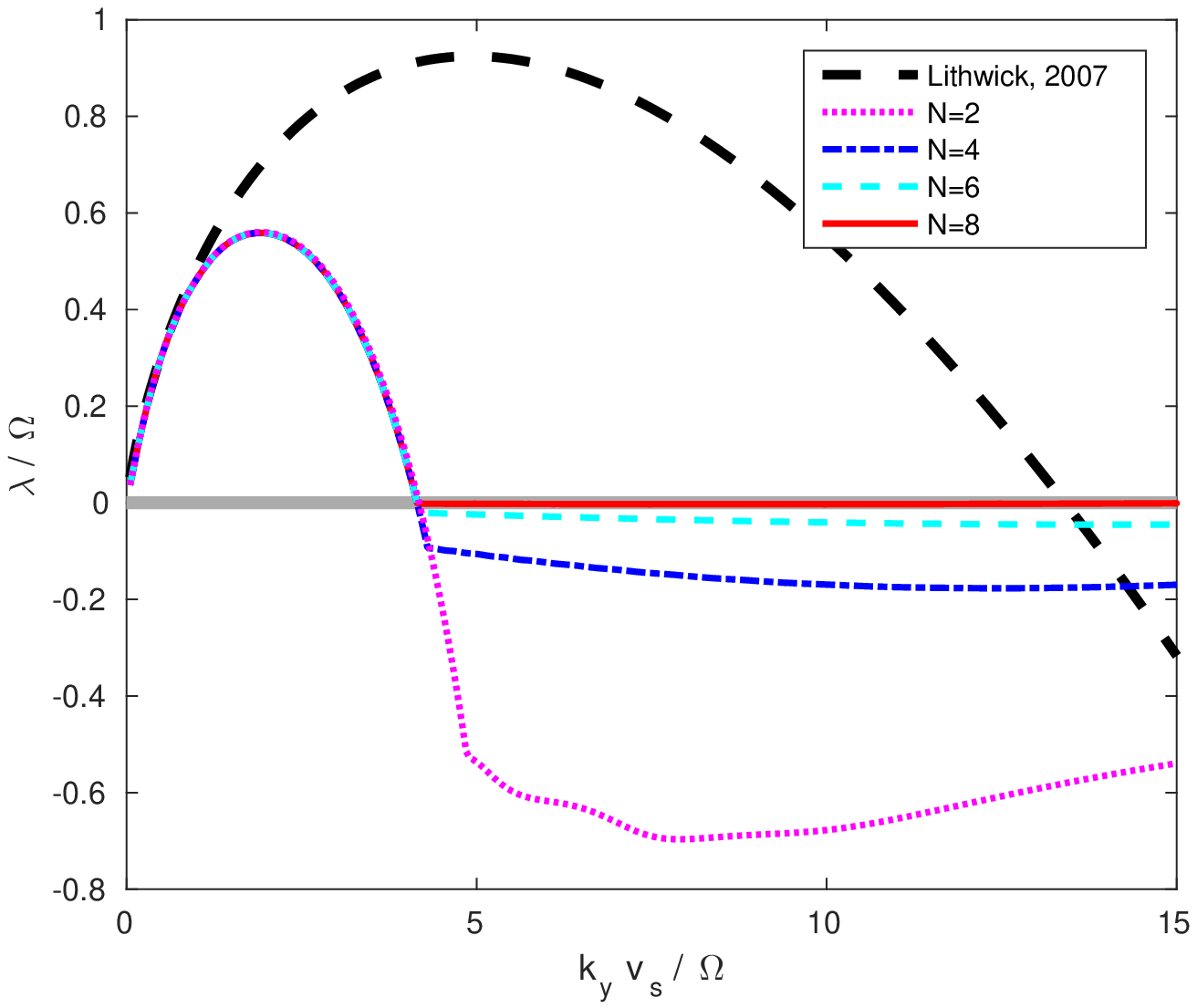}
  \caption{Growth rate $\lambda/\Omega$ as a function of $k_y \varv_s/\Omega$ for $A_h=0.05$, $k \varv_s/\Omega=8$ and a few $N$ values: $N=2$ (magenta, dotted line); $N=4$ (blue, dot-dashed); $N=6$ (cyan, dashed); $N=8$ (red, full). The results are compared to those obtained in \citet{Lithwick2007} (black, thick dashed line) and the zero growth rate line is also plotted as reference. Increasing the number of modes clearly reduces the energy lost from the system, with only 17 modes ($N=8$) required to minimise the loss to a negligible value. This however affects only the decaying modes, as the modes with $\lambda/\Omega>0$ appear unaffected by increasing $N$.}
  \label{fig:growthrate-k8-comparison}
\end{figure}

The first issue to consider is that of energy conservation, for taking into account a finite number of modes inevitably results in some energy being lost from the system, which breaks its time-reversal symmetry. For that purpose, a plot for the growth rate $\lambda/\Omega$ as a function of $k_y \varv_s/\Omega$ for the same parameter values as Figure~\ref{fig:growthrate-k8-N7} was attempted for a few values of $N$ (it's important to remember that the total number of states considered in the calculation is given by $2N+1$); this is shown in Figure \ref{fig:growthrate-k8-comparison}, where the non-conservation of energy is apparent by any negative growth rate values.

The plot shows the growth rate as a function of $k_y \varv_s/\Omega$ for $N=2$ (magenta, dotted line), $N=4$ (blue, dot-dashed), $N=6$ (cyan, dashed) and $N=8$ (red, full) as well as a comparison to the results obtained by \citet{Lithwick2007} and previously discussed in Section~\ref{sec:no-sg}. While the growth rate values of all four runs compare well with the results by \citet{Lithwick2007} (as analysed in more detail in \ref{sec:app-res}), the diminishing negative values for $k_y \varv_s/\Omega \gtrsim 4$ for increasing $N$ show that considering a larger number of states reduces the amount of energy lost from the system, with as little as $N=6$ providing an energy loss small enough to be considered negligible. Increasing $N$ from $N=2$ to $N=8$ does however augment the computational time greatly and while this did not necessarily represent an issue for the non-self-gravitating analysis, computational times were severely lengthened in the self-gravitating case. For that reason it was necessary to minimise numerical resolution, while however making sure that satisfactory results were still obtained.

\subsection{Effects on growth rate}
\label{sec:app-res}
\begin{figure}
	\includegraphics[width=\columnwidth]{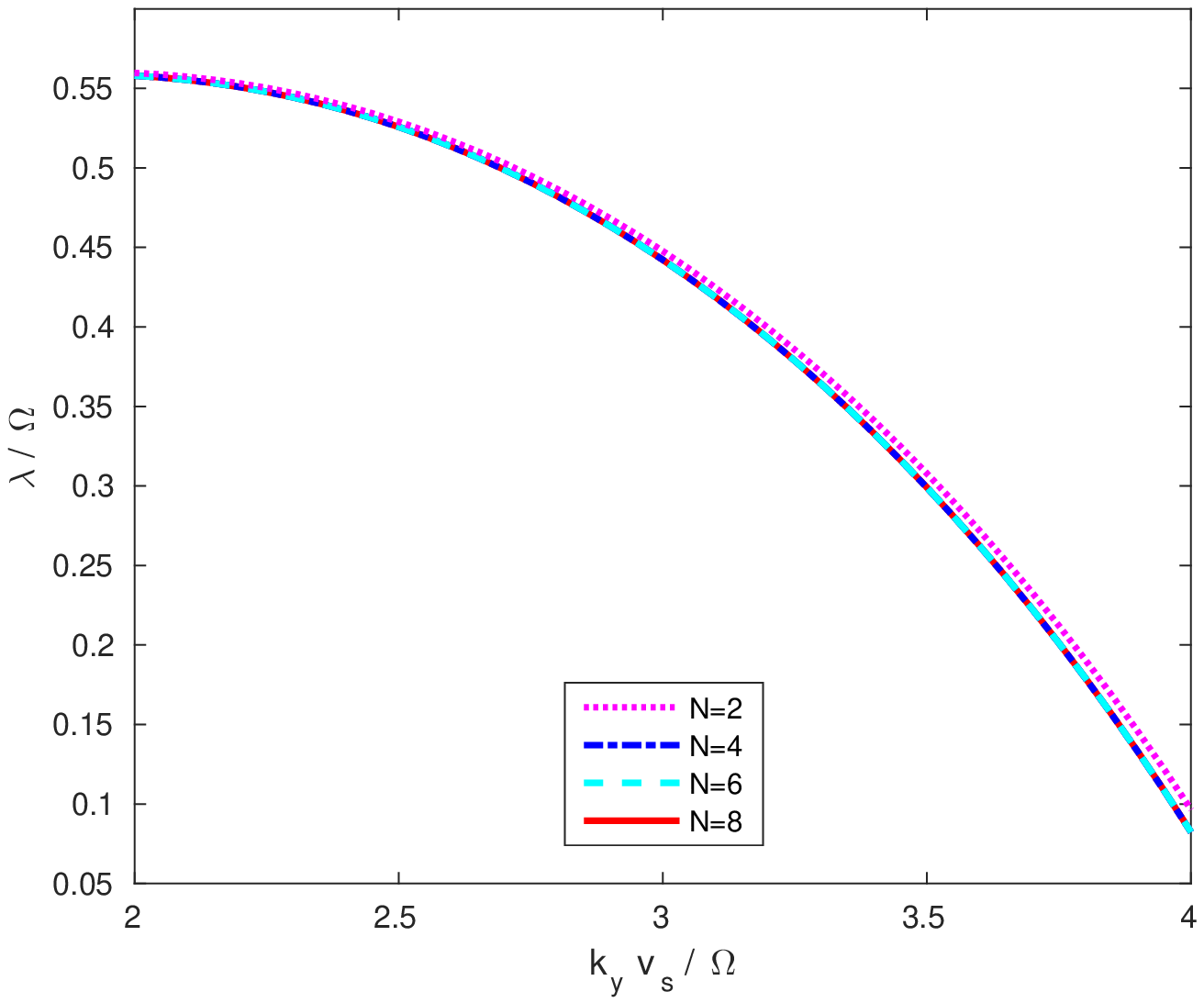}
  \caption{Zoom-in of the growth rate $\lambda$ as a function of $k_y \varv_s/\Omega$ for $A_h=0.05$, $k \varv_s/\Omega=8$ and a few $N$ values: $N=2$ (magenta, dotted line); $N=4$ (blue, dot-dashed); $N=6$ (cyan, dashed); $N=8$ (red, full). The results are compared to those obtained in \citet{Lithwick2007} (black, thick dashed line). Only a small discrepancy visible between the $N=2$ run and the others.}
  \label{fig:growthrate-k8-comparison-zoom}
\end{figure}

\begin{figure}
	\includegraphics[width=\columnwidth]{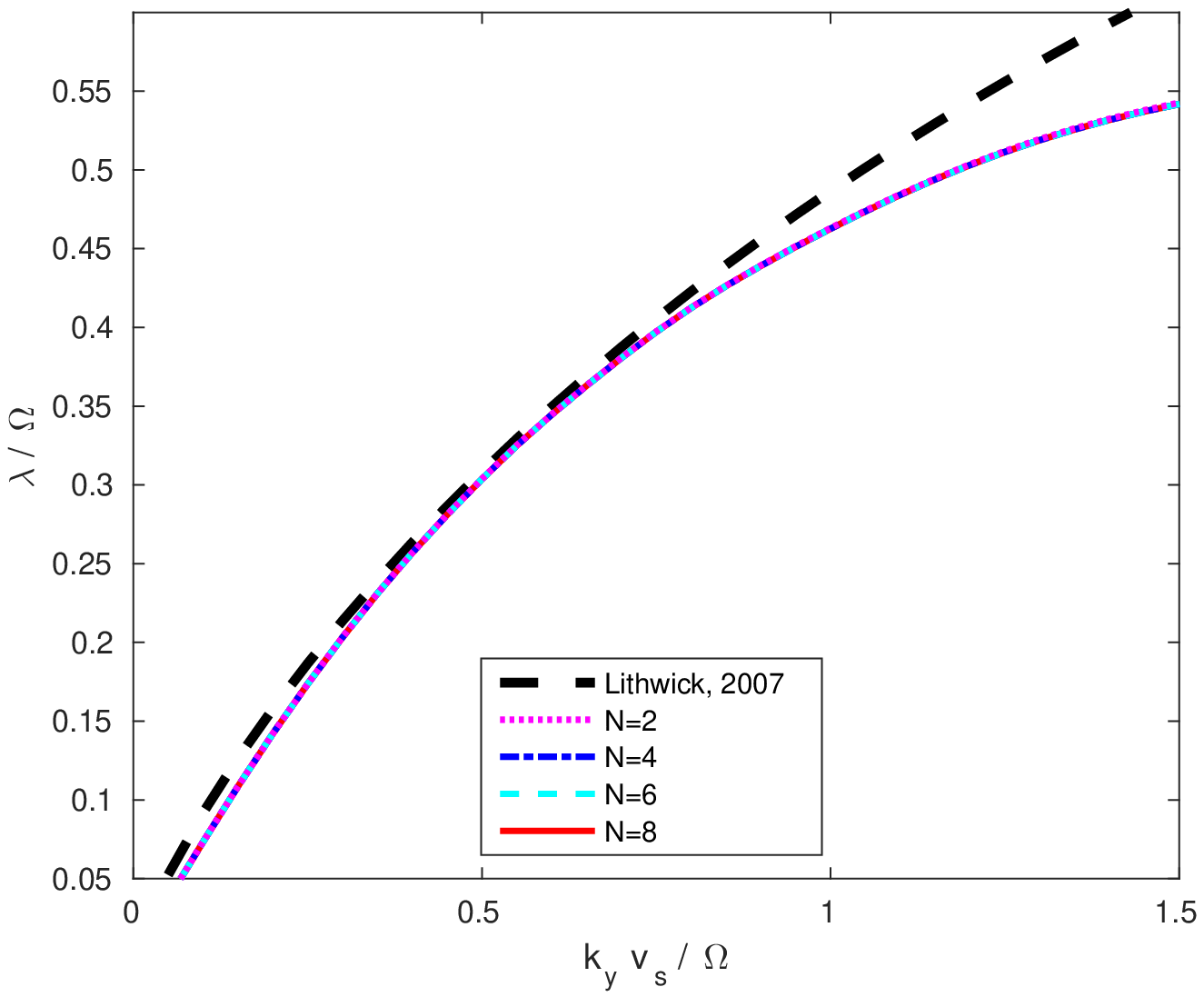}
  \caption{Further zoom-in of the growth rate $\lambda$ as a function of $k_y \varv_s/\Omega$ for $A_h=0.05$, $k \varv_s/\Omega=8$ and a few $N$ values: $N=2$ (magenta, dotted line); $N=4$ (blue, dot-dashed); $N=6$ (cyan, dashed); $N=8$ (red, full). The results are compared to those obtained in \citet{Lithwick2007} (black, thick dashed line). As no discrepancy is visible between the four runs, we conclude that the resolution doesn't impact on the accuracy of the growth rates calculated.}
  \label{fig:growthrate-k8-comparison-zoom2}
\end{figure}

Figure~\ref{fig:growthrate-k8-comparison} can also be analysed in such a way as to determine the impact of numerical resolution upon the accuracy of the growth rate values, simply by focusing on the $k_y \varv_s/\Omega \lesssim 5$ part of the plot. This is done in Figures~\ref{fig:growthrate-k8-comparison-zoom} and ~\ref{fig:growthrate-k8-comparison-zoom2}, which represent zoomed-in versions of Figure~\ref{fig:growthrate-k8-comparison}.

Unlike the previous figure, where increasing $N$ clearly decreased the energy lost by the system, Figure~\ref{fig:growthrate-k8-comparison-zoom} only shows a small discrepancy between the curve for $N=2$ and the others on the descending half of the plot, while the ascending part in Figure~\ref{fig:growthrate-k8-comparison-zoom2} shows no visible discrepancy between the growth rates of the four runs analysed. This very rapid convergence for the instability growth rates allowed us to pick a low $N$ value to minimise the computational time as the energy loss resulting from considering a small number of states was found not to have an impact on the instability growth rate accuracy, which represented the area of interest of this analysis.

\subsection{Undefined neutral modes} \label{sec:undefined-modes}
As shown in Figure~\ref{fig:growthrate-k8-comparison} for $k\varv_s/\Omega=8$, the negative growth rate values quickly converge towards zero as the number of modes $2N+1$ is increased; this is due to a reduction in the amount of energy lost from the system caused by considering more states and hence an extended range of wavenumbers (and wavelengths). 

This convergence was however not present in a lower $k\varv_s/\Omega$ scenarios (eg, $k \varv_s/\Omega=2$), where extremely high values of $N$ were attempted ($N\gtrsim 100$) with no convergence seen. This non-convergence is believed to be caused by the truncation of the ladder of shearing waves, with the decaying modes exhibiting oscillations at the smallest resolved scale as a result.

It seems therefore that only spurious neutral modes may exist for compressible cases (such as $k\varv_s/\Omega=2$) where density waves are present, while we believe regular neutral modes may exist in incompressible conditions (eg. $k\varv_s/\Omega=8$).

However, it is worth stressing that only decaying and neutral modes are affected by the convergence issues mentioned above; these are not part of the analysis presented. Growing modes on the other hand, on which the analysis is based, are fully converged for all values of $N$ looked at in this work for any value of $k \varv_s/\Omega$.




\bsp	
\label{lastpage}
\end{document}